\documentclass[letterpaper,10pt]{article} 

\usepackage{opticameet3} 

\usepackage{amsmath,amssymb}
\usepackage[colorlinks=true,bookmarks=false,citecolor=blue,urlcolor=blue]{hyperref} 
\newcommand*{\affmark}[1][*]{\textsuperscript{#1}}

\begin{document}

\title{Demonstration of Cooperative Transport Interface using open-source 5G OpenRAN and virtualised PON network}


\author{Frank Slyne\affmark[1, *], Kevin O'Sullivan\affmark[2], Merim Dzaferagic\affmark[1], Bruce Richardson\affmark[2], Marcin Wrzeszcz\affmark[2], Brendan Ryan\affmark[2], Niall Power\affmark[2], Robin Giller\affmark[2], Marco Ruffini\affmark[1]}

\address{\affmark[1]CONNECT Centre, School of Computer Science and Statistics, Trinity College Dublin, \affmark[2]Intel Corporation, Ireland}

\email{\{fslyne, mdzafera, marco.ruffini\}@tcd.ie, \{kevin.o.sullivan, bruce.richardson, marcin.wrzeszcz, brendan.ryan,  ,  robin.giller\}@intel.com}


\begin{abstract}
We demonstrate a real-time, converged 5G-PON through the Cooperative Transport Interface, synchronising 5G and PON-DBA upstream schedulers. This innovative approach, implemented using 5G and PON open network implementations, significantly enhances network resource allocation, reducing latency.
\end{abstract}

\section{Overview}

The Open Radio Access Network (O-RAN), an evolution of the Cloud-Radio Access Network (Cloud-RAN) \cite{itu2021}, is transforming network performance by integrating new features and enhancing capacity and latency. The architecture of Cloud-RAN, defined by the Next Generation Mobile Networks (NGMN) alliance, includes the Central Unit (CU), Distributed Unit (DU), and Radio Unit (RU). The connection between these units, known as the Low Layer Split (LLS) \cite{Das2019} transport interface or fronthaul, is traditionally handled by dedicated fiber fronthaul networks. However, replacing point-to-point fibre with Passive Optical Networks (PON), would substantially reduce connectivity costs, enabling a much needed increase in small cell densification, for 5G and beyond. The key issue to be addressed in 5G over PON is the latency generated by the Dynamic Bandwidth Allocation (DBA), which occurs due to lack of coordination between PON and DU upstream schedulers. The concept of "Cooperative DBA" was originally introduced in \cite{Tashiro2014} to address this issue. This focuses on the development of an optical-mobile cooperative interface that translates mobile upstream scheduling data into PON transmit request information. The concept was later standardised as "Cooperative Transport Interface (CTI)" \cite{nomura2017}, which facilitates enhanced coordination between PON and RAN upstream schedulers, leading to more efficient resource allocation and reduced latency. 


In our demonstration, we show the integration of a Passive Optical Network (PON) Dynamic Bandwidth Allocation (DBA) with an open source 5G New Radio (NR) Medium Access Control (MAC) scheduler. The 5G DU and CU are open source from the srsRAN project \cite{srsran}, and the RU is a National Instruments (NI) USRP X310. The transmission was implemented using Tibit PON transceivers, which include a PON MAC layer. In addition, we have modified the PON MAC by integrating it with our virtual DBA technology \cite{vDBA}.  

The key points our demo will address is to show the capability to intercept UE grant requests from the DU, and make use of that information to schedule upstream PON bursts through a software-based virtual DBA, enabling dynamic and flexible bandwidth management. Achieving this required to provide strict synchronisation between PON and RAN schedulers, thereby reducing latency and improving network efficiency. The implementation makes use of Intel networking technologies like AF\_XDP and DPDK for low-latency packet processing and high-performance network packet delivery. It also required to develop mechanisms to access the Tibit OLT architecture, using MongoDB to program grant scheduling within the Tibit hardware.

\section{Innovation}
\begin{figure}[h!]
   \centering
        \includegraphics[width=0.8\linewidth]{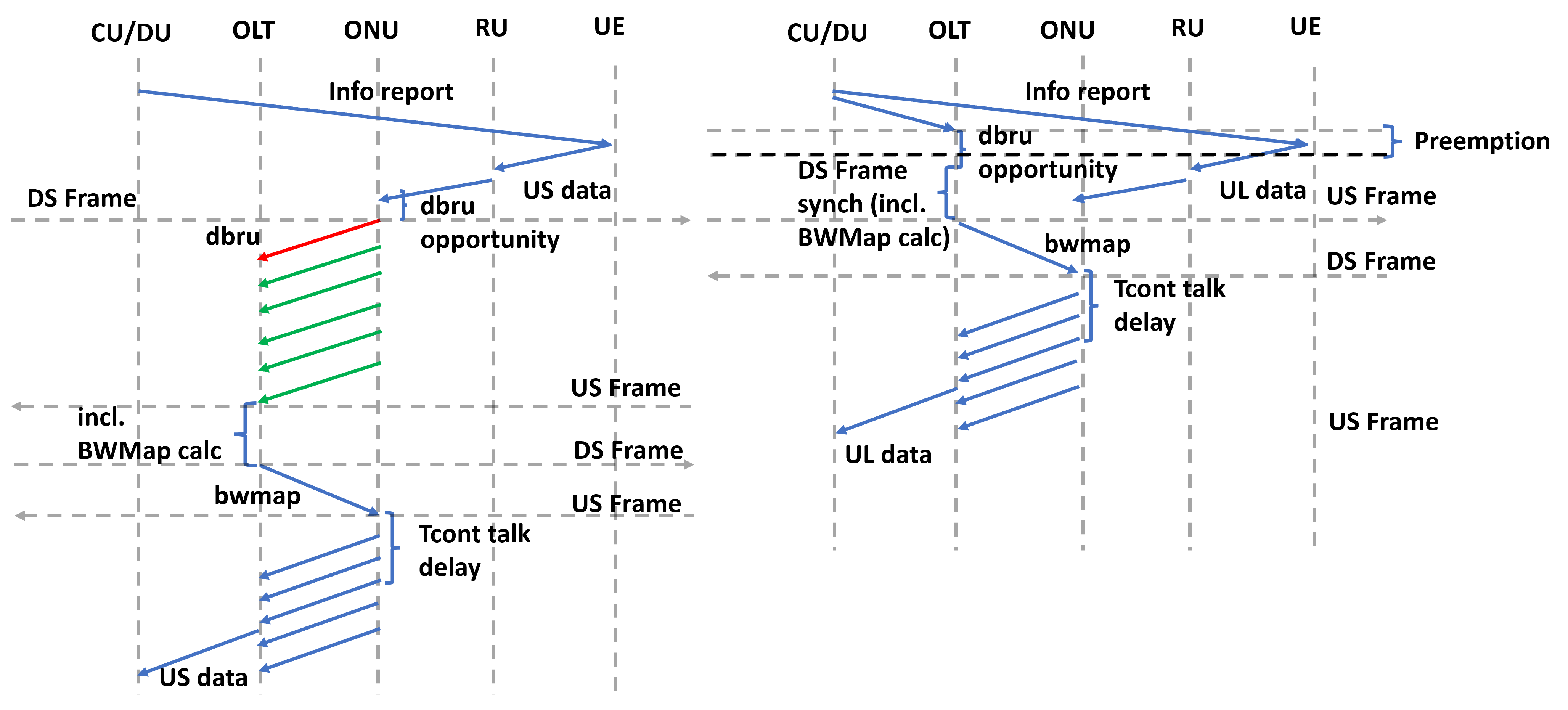}
        \vspace{-4mm}
    \caption{ (a) non-optimised dataflow and (b) optimised dataflow }
    \label{fig:Optimisation}
\end{figure}

The focus of our innovation is the synchronisation of a PON virtual DBA and 5G MAC scheduler using open networking implementations. 
PON and 5G NR systems typically operate independently of each other, which leads to suboptimal resource allocation and data transmission delays affecting QoS and user experience. Unfortunately, this integration can be complicated by the lack of alignment in standards and technologies. To understand the innovation in the synchronisation of MAC scheduling for 5G NR over PON, we consider two sequence diagrams for non-optimised and optimised scheduler scenarios in Fig \ref{fig:Optimisation}. Here, the timeline is downwards, the components (CU/DU, OLT, ONU, RU and UE) through which data flows is left to right. For the non-optimised case (left), a downstream 5G NR frame travels from the CU/DU to the RU over a bearer network (in this case a PON), and over a Radio network to a UE. The PON bearer network is composed of an OLT to ONU fibre network. This 5G NR frame contains an Info Report that signals to the UE to send upstream data. The issue of lack of synchronisation occurs when the ONU receives the 5G NR frame, and stores it in a TCONT buffer in a manner similar to other ONU applications. The frame must wait for an opportunity to transmit upstream, for a duration of time which we call the dbru opportunity time. In the classical implementation, the PON provides no priority to the upstream 5G NR traffic. In the optimised scenario on the right hand side instead, we make use of the UE upstream scheduling information available at the 5G DU, to anticipate the upstream scheduling of the PON DBA.  When the gNodeB (CU/DU) sends the info report to the UE, it also sends an indicator to the PON upstream scheduler to schedule an upstream transmission from  the relevant ONU TCONT buffer, to which the O-RU is connected. This indication is included in the upstream Bandwidth MAP (BWMap) that is included as normal in the downstream frame. 

\vspace{-4mm}
\vspace{-4mm}

To integrate the Tibit OLT DBA with srsRAN, we used Tibit's DBA cascading feature to aggregate upstream scheduling across multiple OLTs within an Ethernet switch or server. We used the Tibit master/slave architecture to software-define the DBA, with one OLT designated as the 'Master' to manage upstream grant scheduling for the entire group. To address the challenge of real-time communication and efficient processing of grant requests, we integrated ZeroMQ (ZMQ) to facilitate the rapid and efficient publishing and reception of grant requests, ensuring low-latency communication between the network components. To ensure coherent operation between the MAC and PHY layers of the srsRAN, we exposed the 5G upstream MAC scheduler grants through the PHY Interface (FAPI). This was crucial for translating MAC layer messages into FAPI messages, enabling effective communication of the scheduler's decisions to the PON network for optimized bandwidth allocation. We encountered a challenge in efficiently managing different types of network traffic, particularly in segregating user-plane and DBA frames. To overcome this, we leveraged the traffic steering capabilities of the Intel X710 NIC. By applying NIC rules for Rx queue steering based on frame type, we were able to separate user-plane and DBA frames into different queues, minimizing buffering time for DBA frames and preventing scheduling delays. Additionally, to achieve fast, low-latency packet processing essential in a telecommunications environment where timing is critical, we implemented AF\_XDP (Address Family - eXpress Data Path). This allowed for high-performance packet delivery directly to user-space programs bypassing most of the Linux networking stack, significantly enhancing data throughput and reducing latency.
\section{OFC Relevance}
Our presentation is particularly relevant to the OFC theme of advanced and virtualised optical access network technologies, including the use of next-generation PONs to facilitate small cell densification in 5G and beyond. The focus is on the demonstration of the convergence of optical access and mobile 5G technologies using open source and open networking technologies, showing an agile, software-driven network management and orchestration. 
This highlights how RAN architecture is evolving to effortlessly connect with optical networks. The Tibit XGS-PON pluggable OLT (Optical Line Terminal) demonstrates open network hardware innovation while providing a small and efficient option for establishing PON networks.

\section{Demo content and Implementation}
\begin{figure}[h!]
   \centering
        \includegraphics[width=0.8\linewidth]{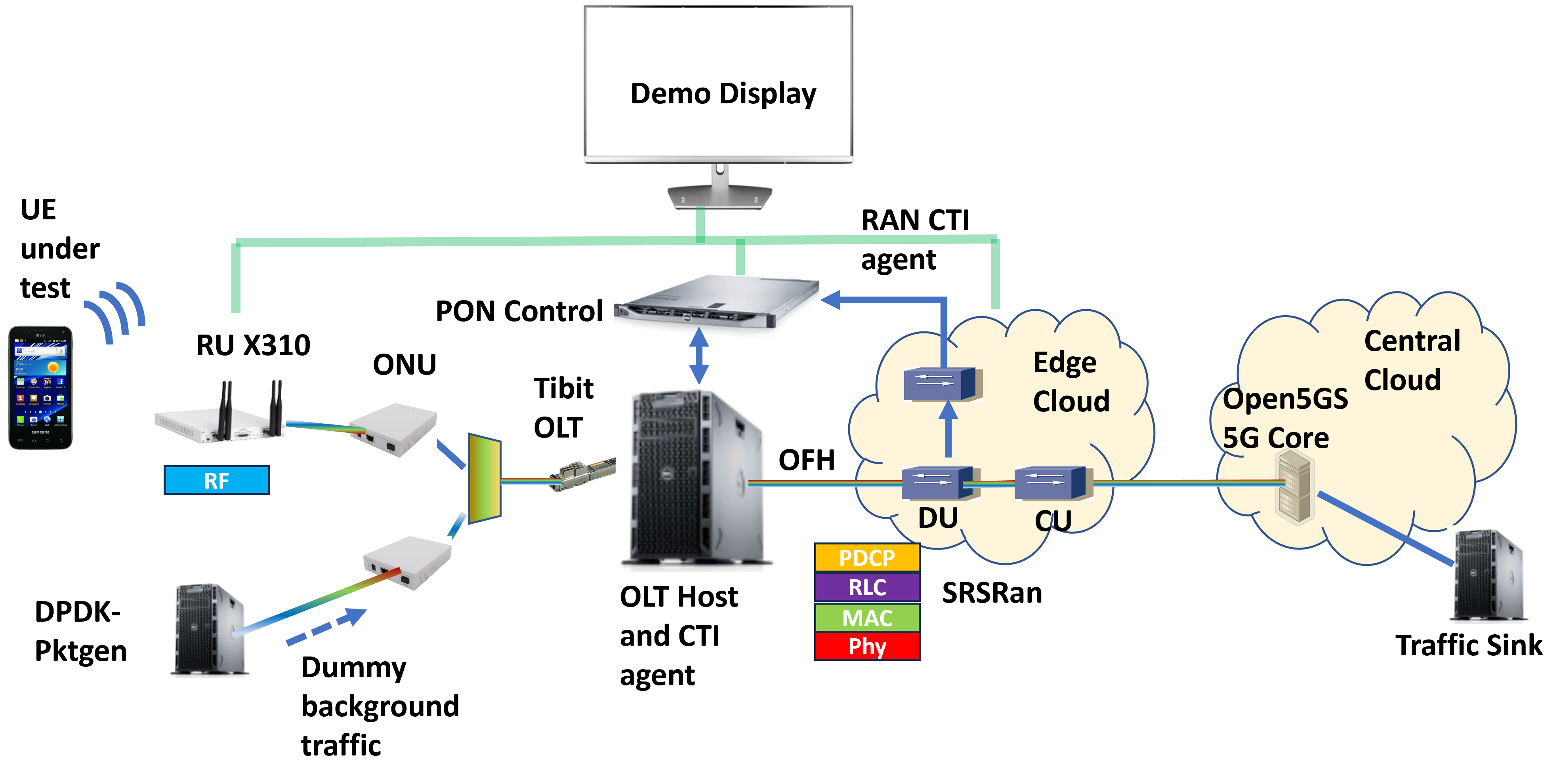}
        \vspace{-4mm}
    \caption{Demonstration Physical Architecture}
    \label{fig:Phys_arch}
\end{figure}

The demonstration will show the integration of Passive Optical Network (PON) virtual Dynamic Bandwidth Allocation (DBA) with 5G NR Medium Access Control (MAC) scheduler. The configuration includes a 5G NR Fronthaul using the srsRAN project. \textbf{The system will be implemented live and fully on site at the demo zone, with no remote elements.} The physical setup will be as shown in Figure \ref{fig:Phys_arch}.

It will consist of hardware USRP X310 (implementing the RU), a PC with a Tibit Micro-pluggable Optical Line Terminal (OLT) (implementing the PHY and MAC of the PON) and servers to implement all other mobile (DU, CU and 5G core) and PON functions (virtual DBA, and other PON software stack).  
For traffic generation, we will use both a UE (commercial smartphone) and a background traffic generator, pushing traffic into the converged 5G/PON system. The purpose of the background traffic generator is to provide contention for bandwidth with the UE device under test, on the PON. It does this by pushing different levels of traffic (50\%, 60\%, 70\%, 80\%, 90\%, 100\%, 110\%) on the PON. We will then show the difference in performance between the uncoordinated system and the synchronised CTI one. In the former, the throughput will be lower and deteriorate as traffic increases, while the latter will maintain the throughput as the PON capacity saturates.

\vspace{-4mm}
\vspace{-4mm}

The demonstration will highlight the decision-making process for resource allocation in scenarios with and without CTI integration, showcasing the efficiency improvements brought by CTI. It will run in real-time, allowing attendees to observe the dynamic interaction between the 5G NR MAC scheduler and PON DBA, mediated by CTI. This will provide a clear view of the differences in network performance, particularly in terms of latency and resource allocation efficiency, under scenarios with and without CTI integration. 
The session will begin with a slide presentation to set the stage for understanding the integration of CTI with PON DBA and 5G NR MAC scheduler. A live data display will offer a side-by-side comparison of network performance for optimized and non-optimized scenarios, with live application-level results and detailed performance measurements

\section{Conclusion and Interaction}
For the demonstration, attendees will first be guided through a presentation explaining the algorithm and operations. As the demo runs in real-time, it will gather and display data on latency and packet arrival, allowing attendees to see live performance metrics. Interaction will be encouraged through Q\&A sessions, offering insights into the system's workings and potential for hands-on engagement with the setup. This approach aims to make the session both informative and engaging.
The demonstration will conclude with a summary of the key observations, emphasising the benefits of integrating CTI in improving network latency and resource allocation efficiency. This will be followed by an interactive Q\&A session, allowing for discussions and questions about the technology and its potential applications. 

\section*{Acknowledgements}
This work was supported by Science Foundation Ireland (SFI) under grants 18/RI/5721 and 13/RC/2077\_p2.

\end{document}